\documentclass[a4paper,11pt]{article}
\pdfoutput=1 

\usepackage{jheppub1} 

\usepackage[T1]{fontenc} 
\usepackage{empheq,mathtools}
\usepackage{booktabs}
\usepackage{xcolor}
\usepackage{verbatim}
\usepackage{float}
\usepackage{subfigure}
\usepackage{framed}
\usepackage{multirow}

\def\tsp{\hspace{0.083333em}}
\def\arXiv#1{\href{http://arxiv.org/abs/#1}{arXiv:#1}}
\def\arXiv#1#2{\href{http://arxiv.org/abs/#1}{arXiv:#1}}
\def\arXivid#1#2{\href{http://arxiv.org/abs/#1/#2}{#1/#2}}
\def\doi#1#2{\href{http://doi.org/#1}{#2}}

\title{\boldmath Holographic supersymmetric R\'enyi entropies from hyperbolic black holes with scalar hair}

\author{Jie Ren}
\author{and Dao-Quan Sun}
\affiliation{School of Physics, Sun Yat-sen University, Guangzhou, 510275, China}

\emailAdd{renjie7@mail.sysu.edu.cn}

\abstract{
We study holographic supersymmetric R\'enyi entropies from a family of hyperbolic black holes in an Einstein-Maxwell-dilaton (EMD) system under the BPS condition. We calculate the thermodynamic quantities of these hyperbolic black holes. We find a remarkably simple formula of the supersymmetric R\'enyi entropy that unifies (interpolates) 11 cases embeddable to 10 or 11 dimensional supergravity. It reproduces many known results in the literature, and gives new results with distinctive features. We show that the supersymmetric version of the modular entropy and the capacity of entanglement cannot be mapped to thermal quantities, due to the dependence of the temperature and the chemical potential by the BPS condition. We also calculate the entanglement spectrum. We derive the potential of the EMD system from a $V=0$ solution and obtain two neutral solutions with scalar hair as a byproduct.
}

\arxivnumber{2404.05638}

\begin{document}

\maketitle

\flushbottom

\section{Introduction}

The R\'{e}nyi entropy is a one-parameter generalization of the entanglement entropy, quantifying the degree of entanglement between two subsystems of a quantum system $A\cup B$:
\begin{equation}
S_{n}=\frac{1}{1-n}\log\text{Tr}[\rho^{n}_{A}],
\end{equation}
where $\rho_A=\text{Tr}_B\rho_{AB}$ is the reduced density operator of $A$. The R\'{e}nyi entropy can be computed by the replica method with the path integral on an $n$-fold cover of the original Euclidean geometry branched around the entanglement surface $\partial A$ \cite{Calabrese:2004eu,Calabrese:2009qy}. The supersymmetric extension of the R\'{e}nyi entropy was introduced in \cite{Nishioka:2013haa} and studied in various dimensions~\cite{Nishioka:2014mwa,Crossley:2014oea,Huang:2014gca,Hama:2014iea,Alday:2014fsa,Huang:2014pda,Zhou:2015cpa,Zhou:2015kaj,Nian:2015xky,Giveon:2015cgs,Mori:2015bro,Nishioka:2016guu,Yankielowicz:2017xkf,Hosseini:2019and}. The supersymmetric R\'{e}nyi entropy is defined as 
\begin{equation}
S^{\text{susy}}_{n}=\frac{1}{1-n}\text{log}\left|\frac{Z^{\text{susy}}_{n}}{(Z^{\text{susy}}_{1})^{n}} \right|,
\end{equation}
where $Z^{\text{susy}}_{n}$ is the supersymmetric partition function on an $n$-branched manifold by turning on an $R$-symmetry background field to preserve supersymmetries. An advantage of the supersymmetric R\'{e}nyi entropy is that it can be calculated using the localization technique~\cite{Pestun:2007rz,Kapustin:2009kz,Jafferis:2010un,Hama:2010av,Hama:2011ea}. By taking the strong coupling and large $N$ limit, it is matched to the holographic calculations by gravity.

The AdS/CFT correspondence offers a convenient way to calculate the R\'enyi entropy for CFTs in the ground state with a spherical entangling surface \cite{Casini:2011kv,Hung:2011nu}. The ordinary R\'enyi entropy can be mapped to the thermal entropy in $\mathbb{S}^1\times\mathbb{H}^{d-1}$ with the temperature $T=1/(2\pi Ln)$, where $L$ is the radius of the hyperbolic space $\mathbb{H}^{d-1}$. In terms of thermal partition functions,
\begin{equation}
Z[\mathbb{S}^d]=Z[\mathbb{S}^1\times\mathbb{H}^{d-1}].
\end{equation}
Similarly, the supersymmetric R\'enyi entropy can be mapped to a twisted thermal partition function. The holographic dual to $\mathbb{S}^1\times\mathbb{H}^{d-1}$ is a hyperbolic black hole (up to a Wick rotation). The supersymmetric R\'enyi entropy can be calculated in terms of charged hyperbolic black holes under the Bogomol'nyi-Prasad-Sommerfield (BPS) condition. The charged R\'enyi entropy is a generalization of the R\'enyi entropy, and it takes into account the distribution of a conserved charge across the entangled states \cite{Belin:2013uta}. While the holographic charged R\'enyi entropy can be calculated by means of charged hyperbolic black holes, the BPS condition imposes a constraint between the temperature and the chemical potential of the black hole.

In this paper, we compute holographic supersymmetric R\'{e}nyi entropy by means of charged hyperbolic black holes in an Einstein-Maxwell-dilaton (EMD) system that interpolates truncations of the following top-down models under the BPS condition.
\begin{itemize}
    \item AdS$_5$/CFT$_4$: $U(1)^3$ truncations of $D=5$ gauged supergravity embeddable to AdS$_5\times \mathbb{S}^5$, near-horizon limit of rotating D3-brane in 10D type IIB supergravity. The CFT dual is $\mathcal{N}=4$ supersymmetric Yang–Mills (SYM) theory \cite{Maldacena:1997re}. Three special cases can be reduced to an EMD system (including when the dilaton vanishes).
    \item AdS$_4$/CFT$_3$: $U(1)^4$ truncations of $D=4$ gauged supergravity embeddable to AdS$_4\times \mathbb{S}^7$, near-horizon limit of rotating M2-brane in 11D supergravity. The CFT dual is the ABJM model \cite{Aharony:2008ug}. Four special cases can be reduced to an EMD system (including when the dilaton vanishes).
    \item AdS$_7$/CFT$_6$: $U(1)^2$ truncations of $D=7$ gauged supergravity embeddable to AdS$_7\times \mathbb{S}^4$, near-horizon limit of rotating M5-brane in 11D supergravity. The CFT dual is 6D $(2,0)$ superconformal field theory. Two special cases can be reduced to an EMD system.
    \item AdS$_6$/CFT$_5$: Romans F(4) gauged supergravity \cite{Romans:1985tw} coupled to matter embeddable to a warped AdS$_6\times \mathbb{S}^4$ background of massive IIA supergravity \cite{Intriligator:1997pq,Cvetic:1999un} or IIB supergravity \cite{Jeong:2013jfc}. Two special cases can be reduced to an EMD system.
\end{itemize}
The metrics of these $D=4,5,6,7$ supergravity theories are summarized in appendix~\ref{sec:STU}. Among these solutions, there are 11 cases of EMD truncations. Interestingly, they can be interpolated by a single EMD system. When the interpolating parameter $\alpha$ takes special values for $AdS_{d+1}$, the EMD system coincides with top-down models. When $\alpha$ does not take these special values, the EMD system in AdS$_4$ still belongs to supergravity. Then we can treat the EMD system in AdS$_4$ as a bottom-up model, since the supersymmetry is gauged in the bulk and global in the boundary.

We find a remarkably simple formula of the supersymmetric R\'enyi entropy that unifies (interpolates) the above 11 special cases; see \eqref{eq:d-sre} below. Among these 11 cases of the supersymmetric R\'enyi entropy, many have appeared in the literature in SCFT and gravity calculations, and some have not appeared in the literature and have new features.

We also calculate the capacity of entanglement \cite{PhysRevLett.105.080501} and the entanglement spectrum \cite{Haldane:2008en}, which are entanglement data expressed in different ways than the R\'enyi entropy. The capacity of entanglement as a quantum information measure different from entanglement entropy has been studied broadly \cite{DeBoer:2018kvc,Nakaguchi:2016zqi,Nakagawa:2017wis,Sun:2023tmv}. In this paper, we generalize the capacity of entanglement to the supersymmetric capacity of entanglement and show that it cannot be mapped to the standard heat capacity of the thermal CFT on hyperbolic space, due to the dependence of the temperature and the chemical potential by the BPS condition. This is different from the previous results of the non-supersymmetric capacity of entanglement. We also calculate the entanglement spectrum, which are eigenvalues of $\rho_A$. We write the entanglement spectrum as convolutions of generalized hypergeometric functions.

At first glance, the dilaton potential in our model looks a little cumbersome. However, we demonstrate that this is the most natural way to add a cosmological constant to the Horowitz-Strominger solution \cite{Horowitz:1991cd} found in 1991. We derive the potential of the EMD system starting with the $V=0$ solution under reasonable assumptions. As a byproduct, we obtain two nontrivial neutral limits as hyperbolic black holes with scalar hair.

The paper is organized as follows. In section~\ref{sec:EMD}, we take advantage of an EMD system to calculate the supersymmetric R\'enyi entropy, and obtain a simple result that interpolates 11 special cases from top-down truncations. In section~\ref{sec:ent-data}, we study the capacity of entanglement and the entanglement spectrum. In section~\ref{sec:gen}, we derive the potential of the scalar field. In section~\ref{sec:sum}, we conclude with some open questions. In appendix~\ref{sec:STU}, we review consistent truncations of $D=10$ and $D=11$ solutions in supergravity. In appendix~\ref{sec:sugra}, we review the FI-gauged supergravity. In appendix~\ref{sec:IR}, we give a special IR geometry.

\section{Supersymmetric R\'{e}nyi entropy from EMD systems}
\label{sec:EMD}
We start with an EMD system that has large intersections with supergravities. In fact, this system in AdS$_4$ or in general dimensions was rediscovered many times \cite{Gao:2004tu,Gao:2004tv,Gao:2005xv,Anabalon:2012ta,Feng:2013tza,Faedo:2015jqa,Ren:2019lgw}. We will give an elegant explanation in section~\ref{sec:Vphi} that the dilaton potential $V(\phi)$ is naturally generated from a $V=0$ solution. The $(d+1)$-dimensional action is
\begin{equation}
S=\int d^{d+1}x\sqrt{-g}\biggl(R-\frac14 e^{-\alpha\phi}F^2-\frac12(\partial\phi)^2-V(\phi)\biggr),\label{eq:actiond}
\end{equation}
where $d\geq 3$, $F=dA$, and $\alpha$ is a parameter. We have set $16\pi G=1$. The potential of the dilaton field is
\begin{equation}
V(\phi)=v_1e^{-\frac{2(d-2)}{(d-1)\alpha}\phi} +v_2e^{\frac{(d-1)\alpha^2-2(d-2)}{2(d-1)\alpha}\phi}+v_3e^{\alpha  \phi }\,,\label{eq:V-alpha}
\end{equation}
where
\begin{equation}
\begin{split}
&v_1=-\frac{(d-1)^2[d(d-1)\alpha^2-2(d-2)^2]\alpha^2}{[2(d-2)+(d-1)\alpha ^2]^2L^2}\,,\\
&v_2=-\frac{8(d-1)^3(d-2)\alpha^2}{{[2(d-2)+(d-1)\alpha^2]^2L^2}}\,,\\
&v_3=-\frac{2(d-1)(d-2)^2[2d-(d-1)\alpha^2]}{[2(d-2)+(d-1)\alpha^2]^2L^2}\,.
\end{split}
\end{equation}
As a justification of this potential, table~\ref{tab:1} below shows that it reproduces 11 EMD systems of especial physical significance. The potential can be written in terms of a superpotential $W(\phi)$ \cite{Lu:2013eoa}:
\begin{align}
V(\phi) &=W'(\phi)^2-\frac{d}{2(d-1)}W^2,\\
W(\phi) &=\frac{2\sqrt{2}\,(d-1)(d-2)}{[2(d-2)+(d-1)\alpha^2]L}\biggl(e^{\frac{\alpha}{2}\phi}+\frac{(d-1)\alpha^2}{2(d-2)}e^{-\frac{d-2}{(d-1)\alpha}\phi}\biggr).
\end{align}
The $\phi\to 0$ behavior of $V(\phi)$ is
\begin{equation}
V(\phi)=-\frac{d(d-1)}{L^2}-\frac{d-2}{L^2}\phi^2+\mathcal{O}(\phi^3)\,,
\end{equation}
where the first term is the cosmological constant, and the second term shows that the mass of the scalar field satisfies $m^2L^2=-2(d-2)$. The scaling dimension of the scalar operator dual to the scalar field $\phi$ satisfies $\Delta(\Delta-d)=m^2L^2$, which has two solutions $\Delta_\pm=2$, $d-2$. Recall that the alternative quantization exists when $-d^2/4\leq m^2\leq -d^2/4+1$; the alternative quantization exists only in $d=3$. The mass is above the BF bound for all $d$ except that the mass saturates the BF bound in $d=4$.

The above EMD system admits an analytic solution \cite{Gao:2004tu,Gao:2004tv,Gao:2005xv}:
\begin{align}
& ds^2 =-f(r)\tsp dt^2+\frac{1}{g(r)}\tsp dr^2+U(r)\tsp d\Sigma_{d-1,\tsp k}^2\,,\label{eq:ansatz-d}\\
& A=2\sqrt{\frac{(d-1)\tsp b\tsp c}{2(d-2)+(d-1)\alpha^2}}\biggl(\frac{1}{r_h^{d-2}}-\frac{1}{r^{d-2}}\biggr)dt\,,\\
& e^{\alpha\phi} =\biggl(1-\frac{b}{r^{d-2}}\biggr)^\frac{2(d-1)\alpha^2}{2(d-2)+(d-1)\alpha^2},
\end{align}
where $d\Sigma_{d-1,\tsp k}^2$ is a $(d-1)$-dimensional hyperbolic space $\mathbb{H}^{d-1}$ ($k=-1$), plane $\mathbb{R}^{d-1}$ ($k=0$), or sphere $\mathbb{S}^{d-1}$ ($k=1$) of unit radius. The functions in the metric are
\begin{align}
f &=\biggl(k-\frac{c}{r^{d-2}}\biggr)\biggl(1-\frac{b}{r^{d-2}}\biggr)^\frac{2(d-2)-(d-1)\alpha^2}{2(d-2)+(d-1)\alpha^2}+\frac{r^2}{L^2}\biggl(1-\frac{b}{r^{d-2}}\biggr)^\frac{2(d-1)\alpha^2}{(d-2)[2(d-2)+(d-1)\alpha^2]},\nonumber\\
g &=f(r)\biggl(1-\frac{b}{r^{d-2}}\biggr)^\frac{2(d-3)(d-1)\alpha^2}{(d-2)[2(d-2)+(d-1)\alpha^2]},\label{eq:soln-d}\\
U &=r^2\biggl(1-\frac{b}{r^{d-2}}\biggr)^\frac{2(d-1)\alpha^2}{(d-2)[2(d-2)+(d-1)\alpha^2]}.\nonumber
\end{align}
The system is invariant under $\alpha\to-\alpha$ and $\phi\to-\phi$; we assume $\alpha\geq 0$. When $\alpha=0$, we obtain the Reissner-N\"ordstrom-AdS (RN-AdS) black hole. Later we take $k=-1$ for hyperbolic black holes. We set $L=1$ in the following.

The mass, temperature, entropy, chemical potential, and charge are given by \cite{Sheykhi:2009pf,Hendi:2010gq}\footnote{For a derivation of the mass by holographic renormalization in AdS$_4$, see appendix A of \cite{Bai:2022obp}.}
\begin{align}
& M=\frac{(d-1)V_\Sigma}{16\pi G}\biggl(c+k\,\frac{2(d-2)-(d-1)\alpha^2}{2(d-2)+(d-1)\alpha^2}\,b\biggr),\label{eq:thermo-Md}\\
& T=\frac{\sqrt{f'g'}}{4\pi}\biggr|_{r=r_h},\qquad S =\frac{V_\Sigma}{4G}U(r_h)^{(d-1)/2}\,,\label{eq:thermo-TSd}\\
& \mu=2\sqrt{\frac{(d-1)\tsp b\tsp c}{2(d-2)+(d-1)\alpha^2}}\frac{1}{r_h^{d-2}}\,,\qquad Q=2(d-2)V_\Sigma\sqrt{\frac{(d-1)\tsp b\tsp c}{2(d-2)+(d-1)\alpha^2}}\,,\label{eq:thermo-muQd}
\end{align}
where $V_\Sigma$ is the volume of $\mathbb{H}^{d-1}$, regulated by integrating out to a maximum radius $R$ of this hyperbolic space \cite{Hung:2011nu}:
\begin{equation}
V_\Sigma\simeq\frac{\Omega_{d-2}}{d-2}\biggl[\frac{R^{d-2}}{\delta^{d-2}}-\frac{(d-2)(d-3)}{2(d-4)}\frac{R^{d-4}}{\delta^{d-4}}+\cdots\biggr],
\end{equation}
where $\Omega_{d-2}=2\pi^{(d-1)/2}/\Gamma((d-2)/2)$ is the area of $\mathbb{S}^{d-2}$. The cutoff $\delta$ is related to the UV cutoff in the dual CFT, consistent with the area law of the entanglement entropy. We have checked that the first law of thermodynamics $dM=TdS+\mu dQ$ is satisfied by~\eqref{eq:thermo-Md}--\eqref{eq:thermo-muQd}. In the grand canonical ensemble, we use the grand potential $\Omega=M-TS-\mu Q$, by which the first law of thermodynamics is
\begin{equation}
d\Omega=-SdT-Qd\mu\,.
\end{equation}
The heat capacity at fixed chemical potential is
\begin {equation}
C_\mu=T\left(\frac{\partial S}{\partial T}\right)_{\mu}=-T\frac{\partial^{2} \Omega(T,\mu)}{\partial T^{2}}\bigg\vert_\mu.
\end{equation}

In the following, we focus on the hyperbolic black holes under the ``BPS condition''
\begin{equation}
c=-b\,.
\label{eq:BPS}
\end{equation}
Recall that the BPS condition for a solution in supergravity is the condition that the Killing spinor equation has nontrivial solutions.\footnote{The technical details to solve the Killing spinor equation can be found in \cite{Romans:1991nq}, for example.} When it is satisfied, the black hole and the corresponding state in the dual CFT preserve a fraction of supersymmetry. This EMD system is related to consistent truncations of supergravity in the following way.
\begin{itemize}
    \item When the parameters $\alpha$ and $d$ take the values in table~\ref{tab:1}, we have checked that \eqref{eq:BPS} is exactly the BPS condition for all 11 cases according to previous studies on $D=4$, $5$, $6$, $7$ supergravity solutions \cite{Huang:2014pda,Zhou:2015kaj,Hosseini:2019and} as consistent truncations of $D=10$ and $D=11$ supergravities. In these cases, the EMD system is a top-down model. 
    \item For AdS$_4$, the system belongs to supergravity for all $\alpha$. The EMD system is obtained by turning off one of two U(1) gauge fields in an FI-gauged supergravity; see appendix~\ref{sec:sugra}. Here \eqref{eq:BPS} is the BPS condition according to \cite{Nozawa:2022upa}. The system is a bottom-up model that interpolates four top-down models.
    \item For AdS$_5$ and higher dimensions, the system may belong to the so-called fake supergravity \cite{Freedman:2003ax,Celi:2004st}, in which we treat \eqref{eq:BPS} as a BPS-like condition.
\end{itemize}

We need to distinguish the extremal limit and the BPS limit for hyperbolic black holes in AdS. For the (asymptotically flat) RN black hole, these two limits coincide. However, in the AdS case, they are not the same.
\begin{itemize}
\item The extremal limit of a black hole is reached when (i) two horizons merge to a degenerate horizon, or (ii) the horizon moves towards the spacetime singularity. In case (i) we obtain an AdS$_2$ factor in the extremal limit. In case (ii) we may obtain hyperscaling-violating geometries.

\item The BPS limit is different from the extremal limit for hyperbolic black holes in AdS. After taking the BPS limit, we can still vary the temperature. For a certain range of $\alpha$, the extremal limit can be taken, and the temperature reaches zero.
\end{itemize}

For the solution \eqref{eq:ansatz-d}--\eqref{eq:soln-d}, the gauge field is imaginary under the BPS condition $c=-b$. Nevertheless, all thermodynamic quantities are well defined, and the gauge field is real in the Euclidean signature, the same as for the hyperbolic RN-AdS black hole \cite{Belin:2013uta}. The curvature singularity is at $r=0$ and $r^{d-2}=b$, and the parameter $b$ can be either positive or negative. The horizon of the black hole is determined by $f(r_h)=0$, from which the parameter $b$ is expressed in terms of $r_h$:
\begin{equation}
b=r_h^{d-2}-r_h^{\frac{(d-2)^2 [2 d-(d-1) \alpha ^2]}{2 (d-2)^2-d (d-1) \alpha ^2}}.\label{eq:brhd}
\end{equation}
The temperature is given by
\begin{equation}
T=
\frac{2(d-1)(d-2)r_h^p-2 (d-2)^2+(d-1)\alpha^2}{2\pi [2 (d-2)+(d-1)\alpha^2]},\label{eq:thermo-T}
\end{equation}
where we have used~\eqref{eq:brhd} to replace $b$ with $r_h$, and
\begin{equation}
p=\frac{(d-2)[2(d-2)+(d-1)\alpha^2]}{2(d-2)^2-(d-1)\alpha^2}.
\end{equation}
From \eqref{eq:thermo-T}, we conclude that the temperature can reach zero only if $0\leq\alpha\leq\alpha_*$, where
\begin{equation}
    \alpha_*=(d-2)\sqrt{\frac{2}{d-1}}\,.
\end{equation}
There are three distinctive classes as follows:
\begin{itemize}
\item $0\leq\alpha<\alpha_*$. The temperature reaches zero when
\begin{equation}
r_h=\biggl(\frac{d-2}{d-1}-\frac{\alpha^2}{2(d-2)}\biggr)^{\frac{1}{2 p}}.
\end{equation}
At zero temperature, the IR geometry is AdS$_2\times\mathbb{H}^{d-1}$. Notice that the $\alpha=0$ case is the RN-AdS$_{d+1}$ black hole.

\item $\alpha=\alpha_*$. The horizon size is $r_h=L$. The temperature is $T=\sqrt{2(1-b)}/4\pi$. The temperature reaches zero when $b=1$. At zero temperature, the IR geometry has a curvature singularity. See appendix~\ref{sec:IR} for the IR geometry.

\item $\alpha>\alpha_*$. The temperature cannot reach zero. There is a minimal temperature at $r_h=0$.
\end{itemize}

After a conformal mapping, the R\'{e}nyi entropy with the entangling surface being a sphere can be calculated from the thermodynamics of the CFT living on $\mathbb{S} \times \mathbb{H}^{d-1}$ \cite{Casini:2011kv,Hung:2011nu}. By the AdS/CFT correspondence, the supersymmetric R\'enyi entropy can be calculated in terms of the hyperbolic black holes under the BPS condition. We have
\begin{equation}
  \text{Tr}[\rho_A^n]=\frac{Z(T_0/n)}{Z(T_0)^n},
\end{equation}
where $Z(T)=\text{Tr}[e^{H/T}]$ is the thermal partition function of the hyperbolic black hole at temperature $T$. For the entangling surface with radius $L$, we have $T_0=1/(2\pi L)$ and $T=T_0/n$. With the grand potential $\Omega=\beta^{-1}I=-T\log Z$ of black holes, where $I$ is the Euclidean on-shell action and $\beta=1/T$, the supersymmetric R\'{e}nyi entropy is given by
\begin{equation}
  S_n=\frac{n}{1-n}\frac{1}{T_0}\bigl[\Omega(T_0,\mu_0)-\Omega(T_0/n,\mu)\bigr].
  \label{eq:Sn-G}
\end{equation}
The integral representation of $S_n$ is \cite{Huang:2014gca}
\begin{equation}
    S_n=\frac{n}{n-1}\int_n^1\partial_{n'}\left(\frac{\log Z(T_0/n',\mu)}{n'}\right)dn'=\frac{n}{n-1}\frac{1}{T_0}\int_{T_0/n}^{T_0}\left(S+Q\frac{d\mu}{dT}\right)dT.
\end{equation}
Note that $T$ and $\mu$ are not independent due to BPS condition.

For our hyperbolic black holes, the grand potential under the condition $c=-b$ is
\begin{equation}
\Omega=M-TS-\mu Q=-\frac{V_\Sigma}{16\pi G}r_h^{\frac{2(d-1)(d-2)^2}{2(d-2)^2-(d-1) \alpha ^2}}\,.
\label{eq:grand}
\end{equation}
The supersymmetric R\'{e}nyi entropies are given by
\begin{equation}
\label{eq:d-sre}
    S_n=\frac{V_\Sigma}{4G}\,\frac{n}{n-1} \Biggl[1-\left(\frac{(d-2)n+1}{(d-1)n}-\frac{(n-1) \alpha^2}{2(d-2)n}\right)^{\frac{2 (d-1) (d-2)}{2 (d-2)+(d-1) \alpha ^2}}\Biggr].
\end{equation}
The entanglement entropy is
\begin{equation}
    S_1=\lim_{n\to 1}S_n=\frac{V_\Sigma}{4G}\quad (\,=4\pi V_\Sigma\,)\,,
\end{equation}
which is the same for all $\alpha$. To have a well-defined $S_n$ for all $n$, we need $\alpha\leq\alpha_*$. For $d=3,4,5,6$, the values of $\alpha_*$ are as follows.
\begin{equation}
\setlength\tabcolsep{8pt}
\begin{tabular}{c|cccc}
& AdS$_4$ & AdS$_5$ & AdS$_6$ & AdS$_7$\\
\hline
\rule{0pt}{12pt}
$\alpha_*$ & 1 & $4/\sqrt{6}$ & $3/\sqrt{2}$ & $8/\sqrt{10}$
\end{tabular}
\end{equation}
As we will see in the next subsection, an inequality of the R\'enyi entropy is always violated when $\alpha>\alpha_*$.

\begin{table}   
\begin{center}   
\caption{\label{tab:1} ``Periodic table'' of top-down supergravity models and special cases of the supersymmetric R\'enyi entropy. The naming convention and the metrics are in appendix~\ref{sec:STU}. (A) The $S_n$ was calculated by the localization method in the SCFT and matches the holographic result. (B) The $S_n$ was calculated for free fields, and matches the holographic result. (C) To our knowledge, no SCFT calculation is known. (D) It violates R\'enyi entropic inequalities.}
\renewcommand{\arraystretch}{2.5}
\setlength\doublerulesep{0.3pt}
\begin{footnotesize}
\begin{tabular}{|@{\hspace{0pt}}c@{\hspace{0pt}}|c|c|c|c|@{\hspace{2pt}}c@{\hspace{2pt}}|}
\hline\hline
  AdS$_{d+1}$ & $\alpha$ & Name & Metric & Supersymmetric R\'enyi entropy & Cases \\
 \hline\hline  \multirow{4}{*}{\renewcommand{\arraystretch}{1.3}\begin{tabular}{c}
     AdS$_4$\\
      ($d=3$)
 \end{tabular}}& $\alpha=0$ & RN-AdS$_4$ & $H_{1,2,3,4}=H$ & $S_n=\dfrac{3n+1}{4n}S_1$ & A \cite{Nishioka:2013haa,Huang:2014gca,Nishioka:2014mwa}\\ \cline{2-6}
           \multirow{4}{*}{ }    & $\alpha=1/\sqrt{3}$ & 3-charge & $H_{1,2,3}=H$, $H_4=1$ & $S_n=\dfrac{n}{n-1}\biggl[1-\biggl(\dfrac{n+2}{3n}\biggr)^{3/2}\biggr]S_1$ & C \\ \cline{2-6}
   \multirow{4}{*}{ } & $\alpha=1$ & 2-charge & $H_{1,2}=H$, $H_{3,4}=1$ & $S_n=S_1$ & C \\  \cline{2-6}
    \multirow{4}{*}{ } & $\alpha=\sqrt{3}$ & 1-charge & $H_1=H$, $H_{2,3,4}=1$ & $S_n=\dfrac{n}{n-1}\biggl[1-\biggl(\dfrac{2-n}{n}\biggr)^{1/2}\biggr]S_1$ & C, D\\
 \hline\hline
  \multirow{3}{*}{\renewcommand{\arraystretch}{1.3}\begin{tabular}{c}
       AdS$_5$\\
       ($d=4$)
  \end{tabular}}& $\alpha=0$ & RN-AdS$_5$ & $H_{1,2,3}=H$ & $S_n=\dfrac{19n^2+7n+1}{27n^2}S_1$ & B \cite{Huang:2014pda,Zhou:2015cpa}\\ \cline{2-6}
   \multirow{3}{*}{ }    & $\alpha=2/\sqrt{6}$ & 2-charge & $H_{1,2}=H$, $H_3=1$ & $S_n=\dfrac{3n+1}{4n}S_1$ & A, B \cite{Huang:2014pda} \\ \cline{2-6}
   \multirow{3}{*}{ } & $\alpha=4/\sqrt{6}$ & 1-charge & $H_1=H$, $H_{2,3}=1$ & $S_n=S_1$ & B \cite{Huang:2014pda} \\ 
  \hline\hline
\multirow{2}{*}{\renewcommand{\arraystretch}{1.3}\begin{tabular}{c}
       AdS$_6$\\
       ($d=5$)
  \end{tabular}}& $\alpha=1/\sqrt{2}$ & 2-charge & $H_1=H_2=H$ & $S_n=\dfrac{19n^2+7n+1}{27n^2}S_1$ & A \cite{Hosseini:2019and,Hama:2014iea,Alday:2014fsa} \\ \cline{2-6}
   \multirow{2}{*}{ }    & $\alpha=5/\sqrt{10}$ & 1-charge & $H_1=H$, $H_2=1$ & $S_n=\dfrac{n}{n-1}\biggl[1-\biggl(\dfrac{n+2}{3n}\biggr)^{3/2}\biggr]S_1$ & C \\
\hline\hline
\multirow{2}{*}{\renewcommand{\arraystretch}{1.3}\begin{tabular}{c}
       AdS$_7$\\
       ($d=6$)
  \end{tabular}}& $\alpha=2/\sqrt{10}$ & 2-charge & $H_1=H_2=H$ & $S_n=\dfrac{175n^3+67n^2+13n+1}{256n}S_1$ & B \cite{Zhou:2015kaj} \\ \cline{2-6}
   \multirow{2}{*}{ }    & $\alpha=6/\sqrt{15}$ & 1-charge & $H_1=H$, $H_2=1$ & $S_n=\dfrac{3n+1}{4n}S_1$ & B \cite{Zhou:2015kaj} \\
\hline\hline
\end{tabular}
\end{footnotesize}
\end{center}
\end{table}

For 11 special cases belonging to top-down models, we list them with their supersymmetric R\'enyi entropies in table~\ref{tab:1}. For AdS$_4$ and AdS$_5$, $\alpha=0$ gives the RN-AdS black hole. We observe distinctive features as follows. The R\'enyi entropies reproduce known results (A) and (B) while giving new features (C) and (D).
\begin{itemize}
    \item[(A)] RN-AdS$_4$, 2-charge black hole in AdS$_5$, 2-charge black hole in AdS$_6$: The SCFT calculations of $S_n$ have been performed by the localization technique and they match the holographic result \cite{Nishioka:2013haa,Huang:2014gca,Nishioka:2014mwa,Huang:2014pda,Hosseini:2019and,Hama:2014iea,Alday:2014fsa}.
    \item[(B)] All cases in AdS$_5$ and AdS$_7$: The $S_n$ was calculated by the heat kernel method for free fields, and they match the holographic result \cite{Huang:2014pda,Zhou:2015cpa,Zhou:2015kaj}.
    \item[(C)] 1-, 2-, and 3-charge black holes in AdS$_4$ and 1-charge black hole in AdS$_6$: These cases have not been compared with SCFT calculations. The holographic result can be obtained by special cases of \cite{Hosseini:2019and}.\footnote{In a private communication, we learned that Yang Zhou has obtained the $S_n$ for the 1-, 2-, and 3-charge black holes in AdS$_4$, though these findings were not published.}
    \item[(D)] 1-charge black hole in AdS$_4$ is peculiar. When $d=3$, $\alpha=\sqrt{3}$, we have $r_h=\sqrt{n/(2-n)}$, which is real only when $0<n<2$. This is the only case in which $\alpha>\alpha_*$ among the 11 cases.
\end{itemize}

As a remark, the BPS condition~\eqref{eq:BPS} significantly simplifies the thermodynamic quantities. If the parameters $b$ and $c$ are arbitrary, no explicit solution is available for the R\'enyi entropy $S_n$. Another condition in which $S_n$ is explicitly solvable is $c=0$, which was studied in detail in~\cite{Bai:2022obp}, based on a nontrivial neutral limit of these black holes \cite{Ren:2019lgw}.

\section{Exporing the entanglement data}
\label{sec:ent-data}
\subsection{Modular entropy and capacity of entanglement}
\label{sec:CoE}
The R\'enyi entropy as an information-theoretic quantity is related to the thermal entropy on $\mathbb{S}^1\times\mathbb{H}^{d-1}$. The inequalities for the R\'enyi entropy have been proposed in quantum information \cite{Beck:1993} and studied holographically \cite{Hung:2011nu}:
\begin{align}
\frac{\partial}{\partial n}\left(\frac{n-1}{n}S_{n}\right) &\geq 0,\\
\frac{\partial^{2}}{\partial n^{2}}\left((n-1)S_{n}\right) &\leq 0.
\end{align}
The first one corresponds to the positivity of the modular entropy, and the second one corresponds to the positivity of the specific heat. The modular entropy has a geometric interpretation \cite{Dong:2016fnf}. The capacity of entanglement as an important measure of quantum information was originally introduced in \cite{PhysRevLett.105.080501}, and then studied in holography \cite{Nakaguchi:2016zqi}. For the supersymmetric R\'enyi entropy calculated by hyperbolic holes, we find that these inequalities are not satisfied when $\alpha>\alpha_*$.

In terms of the supersymmetric R\'enyi entropy, we define the supersymmetric modular entropy as
\begin{equation}
    \widetilde{S}^\text{susy}_n=n^2\partial_n\Bigl(\frac{n-1}{n}S_n^\text{susy}\Bigr),\label{eq:Stn-def}
\end{equation}
and the supersymmetric capacity of entanglement as
\begin{equation}
C_{E}^{\text{susy}}(n)=n^{2} \partial_{n}^{2}[(1-n)S_{n}^{\text{susy}}].\label{eq:CE-def}
\end{equation}
When $n \to 1$, $C_{E}^{\text{susy}}(1)=\langle (-\text{log}\rho_{A})^{2}\rangle -\langle -\text{log}\rho_{A}\rangle^{2}$ gives the quantum fluctuation with respect to the original state $\rho_{A}$.


We find that the modular entropy $\widetilde{S}_n^\text{susy}$ no longer equals the thermal entropy of the hyperbolic black hole due to the fact that the BPS condition puts a constraint on the temperature and the chemical potential. Similarly, the capacity of entanglement $C_{E}^{\text{susy}}(n)$ cannot map to the heat capacity $C_\mu$ of the thermal CFT on hyperbolic space, unlike the non-supersymmetric capacity of entanglement. More precisely,
\begin{align}
    \widetilde{S}_n^\text{susy} &=S+Q\frac{d\mu}{dT}\\
    C_{E}^{\text{susy}}(n) &=C_\mu+3T\tsp\frac{dQ}{dT}\tsp\frac{d\mu}{dT}+TQ\tsp\frac{d^2\mu}{dT^2}.
\end{align}
The latter was obtained as follows. From \eqref{eq:Sn-G} and \eqref{eq:CE-def}, by $T=T_{0}/n$, we obtain
\begin{equation}
\begin{aligned}
\label{eq:ce3}
C_{E}^{\text{susy}}(n)&=-\frac{d}{dT}\biggl[T^2\frac{d}{dT}\biggl(\frac{1}{T}\Omega(T,\mu)\biggr)\biggr]=-T\frac{d^2}{dT^2}\Omega(T,\mu)\\
&=C_\mu-2T\tsp\frac{\partial^{2}\Omega(T,\mu)}{\partial T \partial \mu}\tsp\frac{d\mu}{dT}-T\tsp\frac{\partial^{2}\Omega(T,\mu)}{\partial \mu^2}\biggl(\frac{d\mu}{dT}\biggr)^2-T\tsp\frac{\partial \Omega(T,\mu) }{\partial \mu}\tsp\frac{d^2\mu}{dT^2},
\end{aligned}
\end {equation}
where the chemical potential $\mu$ and the charge $Q$ depend on the temperature $T$ for supersymmetric states.

From \eqref{eq:d-sre} and \eqref{eq:CE-def}, we obtain the capacity of entanglement
\begin {equation}
\label{eq:cn-2}
C_{E}(n)=\frac{1}{n}\left(\frac{(d-2)n+1}{(d-1)n}-\frac{(n-1) \alpha^2}{2(d-2)n}\right)^{\frac{2 (d-1) (d-2)}{2 (d-2)+(d-1) \alpha ^2}-2}C_E(1),
\end {equation}
where
\begin{equation}
\label{eq:ce2}
C_{E}(1)=\biggl(\frac{d-2}{d-1}-\frac{\alpha^2}{2(d-2)}\biggr)S_{1}.
\end{equation}
For 11 special cases belonging to top-down models, we list their supersymmetric capacity of entanglement in table~\ref{tab:2}. The ratio $C_{E}(1)/S_{1}$ of the capacity of entanglement yields universal information characterizing the dual CFTs~\cite{DeBoer:2018kvc}. From above result, we can see that the ratio $C_{E}(1)/S_{1}$ of the supersymmetric capacity of entanglement is different from the ratio $C_{E}(1)/S_{1}=1$ of the non-supersymmetric capacity of entanglement for neutral black holes~\cite{DeBoer:2018kvc,Sun:2023tmv}.

We can define a heat capacity for black holes under the BPS condition
\begin {equation}
C_\text{BPS}=T\frac{dS}{dT}=\frac{1}{n}\left(\frac{(d-2)n+1}{(d-1)n}-\frac{(n-1) \alpha^2}{2(d-2)n}\right)^{\frac{2 (d-1) (d-2)}{2 (d-2)+(d-1) \alpha ^2}-1}S_1\,.
\end{equation}
It can be verified that the above result equals $(1/n)\widetilde{S}_n$, where $\widetilde{S}_n$ is calculated by \eqref{eq:Stn-def}.

\begin{table}   
\begin{center}   
\caption{\label{tab:2} Special cases of the supersymmetric entanglement of capacity.}
\renewcommand{\arraystretch}{2.5}
\setlength\doublerulesep{0.3pt}
\begin{footnotesize}
\begin{tabular}{|c|c|c|c|c|}
\hline\hline
  AdS$_{d+1}$ & $\alpha$ & Name & $S_n$ & $\quad C_E(t)\quad (\tsp t\equiv 1/n=2\pi T$\tsp) \\
 \hline\hline  \multirow{4}{*}{\renewcommand{\arraystretch}{1.3}\begin{tabular}{c}
     AdS$_4$\\
      ($d=3$)
 \end{tabular}}& $\alpha=0$ & RN-AdS$_4$ & $\dfrac{3n+1}{4n}S_1$ & $C_E(t)=\dfrac{1}{2}tS_1$\\ \cline{2-5}
           \multirow{4}{*}{ }    & $\alpha=1/\sqrt{3}$ & 3-charge & $S_n=\dfrac{n}{n-1}\biggl[1-\biggl(\dfrac{n+2}{3n}\biggr)^{3/2}\biggr]S_1$ & $C_E(t)=t(3+6t)^{-1/2}S_1$ \\ \cline{2-5}
   \multirow{4}{*}{ } & $\alpha=1$ & 2-charge & $S_n=S_1$ &  $C_E(t)=0$\\  \cline{2-5}
    \multirow{4}{*}{ } & $\alpha=\sqrt{3}$ & 1-charge & $S_n=\dfrac{n}{n-1}\biggl[1-\biggl(\dfrac{2-n}{n}\biggr)^{1/2}\biggr]S_1$ & $C_E(t)=-t(-1+2t)^{-3/2}S_1$\\
 \hline\hline
  \multirow{3}{*}{\renewcommand{\arraystretch}{1.3}\begin{tabular}{c}
       AdS$_5$\\
       ($d=4$)
  \end{tabular}}& $\alpha=0$ & RN-AdS$_5$ & $S_n=\dfrac{19n^2+7n+1}{27n^2}S_1$ & $C_E(t)=\dfrac{2}{9}t(2+t)S_1$\\ \cline{2-5}
   \multirow{3}{*}{ }    & $\alpha=2/\sqrt{6}$ & 2-charge & $S_n=\dfrac{3n+1}{4n}S_1$ & $C_E(t)=\dfrac{1}{2}tS_1$\\ \cline{2-5}
   \multirow{3}{*}{ } & $\alpha=4/\sqrt{6}$ & 1-charge & $S_n=S_1$ & $C_E(t)=0$\\ 
  \hline\hline
\multirow{2}{*}{\renewcommand{\arraystretch}{1.3}\begin{tabular}{c}
       AdS$_6$\\
       ($d=5$)
  \end{tabular}}& $\alpha=1/\sqrt{2}$ & 2-charge & $S_n=\dfrac{19n^2+7n+1}{27n^2}S_1$ & $C_E(t)=\dfrac{2}{9}t(2+t)S_1$ \\ \cline{2-5}
   \multirow{2}{*}{ }    & $\alpha=5/\sqrt{10}$ & 1-charge & $S_n=\dfrac{n}{n-1}\biggl[1-\biggl(\dfrac{n+2}{3n}\biggr)^{3/2}\biggr]S_1$ & $C_E(t)=t(3+6t)^{-1/2}S_1$\\
\hline\hline
\multirow{2}{*}{\renewcommand{\arraystretch}{1.3}\begin{tabular}{c}
       AdS$_7$\\
       ($d=6$)
  \end{tabular}}& $\alpha=2/\sqrt{10}$ & 2-charge & $S_n=\dfrac{175n^3+67n^2+13n+1}{256n}S_1$ & $C_E(t)=\dfrac{3}{64}t(3+t)^2S_1$ \\ \cline{2-5}
   \multirow{2}{*}{ }    & $\alpha=6/\sqrt{15}$ & 1-charge & $S_n=\dfrac{3n+1}{4n}S_1$ & $C_E(t)=\dfrac{1}{2}tS_1$\\
\hline\hline
\end{tabular}
\end{footnotesize}
\end{center}
\end{table}

\subsection{Entanglement spectrum}
\label{sec:spectrum}
The R\'{e}nyi entropy $S_n$ for all $n$ determines the entanglement spectrum, which is the eigenvalue distribution of the reduced density matrix $\rho_A$. The holographic result of $S_n$ obtained in the last section is analytic at $n=\infty$. Assuming this analyticity, the entanglement spectrum must include both discrete and continuous parts, with one discrete eigenvalue $\lambda_1$ being the largest eigenvalue of the continuous spectrum \cite{Hung:2011nu}. Thus, the R\'{e}nyi entropy can be written as 
\begin{equation}
  S_n=\frac{1}{1-n} \log \text{Tr} [\rho^n]=\frac{1}{1-n}\log \biggl[d_1 \lambda_1^n+\int_{0}^{\lambda_1}{\bar{\rho}(\lambda) \lambda^n d\lambda}\biggr],
  \label{eq:R\'enyi-eigen}
\end{equation}
where $\bar{\rho}(\lambda)$ is the continuous part of the entanglement spectrum $\rho(\lambda)$. By writing the discrete part into $\rho(\lambda)$ via a Dirac delta function, the R\'{e}nyi entropy satisfies
\begin{equation}
  e^{(1-n)S_n}=\int_{t_1}^{+\infty}{e^{-(n+1)t}\rho(e^{-{t}})dt},
  \label{eq:Laplace}
\end{equation}
where $\lambda$ is reparameterized as $\lambda=e^{-t}$, and $\lambda_1=e^{-t_1}$. This is essentially a Laplace transform with $n$ being the parameter. Thus, the spectrum can be obtained from an inverse Laplace transform,
\begin{equation}
\rho(\lambda)=\frac{1}{\lambda}\mathcal{L}^{-1}\bigl[e^{(1-n)S_n},n,t \bigr] \bigl|_{t=-\log \lambda}=\frac{1}{\lambda}\frac{1}{2\pi i}\lim_{T\rightarrow \infty}\int_{\gamma-iT}^{\gamma+iT}{e^{(1-n)S_n}e^{nt}dn},
\label{eq:invLaplace}
\end{equation}
where the integral is taken over a vertical line with $\text{Re}(s)=\gamma$, and $\gamma$ is a real number ensuring no singularity on the right side of this line.

Assuming that the R\'{e}nyi entropy can be expanded near $n=\infty$ as 
\begin{equation}
  S_{n}=\sum_{i=0}^\infty s_in^{-i}=s_0+\frac{s_1}{n}+\frac{s_2}{n^2}+\cdots,
  \label{eq:R\'enyi-expand}
\end{equation}
where the constant term is related to the largest eigenvalue of the spectrum by $\lambda_1=e^{-s_0}$. For the R\'enyi entropy given by \eqref{eq:d-sre}, the coefficients of the first two terms are
\begin{align}
  \begin{split}
    s_0&=\frac{V_\Sigma}{4G}\Biggl[1-\left(\frac{d-2}{d-1}-\frac{\alpha^2}{2(d-2)}\right)^{\frac{2 (d-1) (d-2)}{2 (d-2)+(d-1) \alpha ^2}}\Biggr],\\
    s_1&=s_0-\frac{V_\Sigma}{4G}\,\frac{2(d-1)(d-2)}{2(d-2)+(d-1)\alpha^2}\left(\frac{d-2}{d-1}-\frac{\alpha^2}{2(d-2)}\right)^{\frac{2 (d-1) (d-2)}{2 (d-2)+(d-1) \alpha ^2}}.
  \end{split}
\end{align}
From table~\ref{tab:1}, we can see that the series terminates at finite orders of $n^{-1}$ in many cases.

In the following, we will give a way to express the entanglement spectrum, which is the inverse Laplace transform of
\begin{equation}
    e^{(1-n)S_n}=e^{s_0-s_1}e^{-s_0n}\exp \biggl(\sum\limits_{i=1}^{\infty}{u_i n^{-i}}\biggr),\qquad u_i=s_i-s_{i+1}.
\end{equation}
By the convolution theorem, we obtain
\begin{equation}
    \rho(\lambda)=\frac{e^{s_0-s_1}}{\lambda}\mathcal{L}^{-1}[e^{-s_0n}]*\mathcal{L}^{-1}[e^{u_1n^{-1}}]*\mathcal{L}^{-1}[e^{u_2n^{-2}}]*\cdots
\end{equation}
The convolution satisfies commutativity and associativity. The inverse Laplace transform can be done term by term. The results are
\begin{align}
    \frac{1}{\lambda}\tsp\mathcal{L}^{-1}[e^{-s_0n}] &=\frac{1}{\lambda}\tsp\delta(t-s_0)=\delta(\lambda-\lambda_1),\\
    \mathcal{L}^{-1}[e^{-s_in^{-i}}] &=\delta(t)+\frac{u_it^{i-1}}{(i-1)!}\,{_0F_i}\biggl(;1+\frac{1}{i},1+\frac{2}{i},\cdots,2;\frac{u_it^i}{i^i}\biggr).
\end{align}
where ${_0F_i}$ is a generalized hypergeometric function. Generally, it is unlikely to write a closed form for the above convolutions. If the large-$n$ expansion of $S_n$ terminates as $S_n=s_0+s_1/n$, which happens three times in table~\ref{tab:1}, a closed form expression can be obtained in the following. The function $_0F_1$ can be expressed in terms of the modified Bessel function of the first kind:
\begin{equation}
    _0F_1(;2;z)=\frac{1}{\sqrt{z}}I_1(2\sqrt{z})\,.
\end{equation}
As a consequence, the entanglement spectrum for $S_n=s_0+s_1/n$ is
\begin{equation}
\rho(\lambda)=e^{s_0-s_1}\delta(\lambda_1-\lambda)+\frac{s_1\theta(\lambda_1-\lambda)}{\lambda \sqrt{s_1\ln(\lambda_1/\lambda)}}I_1\bigl(2\sqrt{s_1\ln(\lambda_1/\lambda)}\bigr),
\label{eq:spectrum-I1}
\end{equation}
where $\theta(x)$ is the Heaviside step function. When $s_0=s_1$, this reproduces the well-known result of the entanglement spectrum for 2D CFTs \cite{Calabrese:2008cl}.

As a comparison, the entanglement spectrum is expressed in terms an infinite sum \cite{Bai:2022obp} (by the approach in \cite{Belin:2013dva})
\begin{equation}
    \rho(\lambda)
    =e^{s_0-s_1}\biggl(\delta(\lambda_1-\lambda)+\frac{\theta(\lambda_1-\lambda)}{\lambda}\sum_{i=0}^{\infty}{\frac{v_{i+1}}{i!} \bigl(\ln(\lambda_1/\lambda)\bigr)^{i}}\biggr),
    \label{eq:rho-vi}
\end{equation}
where the coefficients $\{v_i\}$ can be calculated order by order by expanding
\begin{equation}
\exp \biggl(\sum\limits_{i=1}^{\infty}{u_i n^{-i}}\biggr)=1+\sum\limits_{i=1}^{\infty}{v_i n^{-i}}.
\end{equation}
The sum~\eqref{eq:rho-vi} captures the low-lying part of the spectrum \cite{Bai:2022obp}, where ``low-lying'' represents the spectrum with lower energy, or with $\lambda$ closer to the largest eigenvalue \cite{Haldane:2008en}.

For $\lambda \to 0$, the spectrum can be approximated by the saddle point method. The saddle point $n_0$ is given by 
\begin{equation}
  \frac{\partial}{\partial n}\bigl((n-1)S_n\bigr) \biggl|_{n_0}+\ln{\lambda}=0,
  \label{eq:saddle-point}
\end{equation}
and the integral \eqref{eq:invLaplace} is approximated by
\begin{equation}
\label{eq:ce-saddle-point}
  \rho(\lambda \to 0)\sim \frac{1}{\lambda^{n+1}} e^{(1-n)S_{n}} \Bigl[ 2\pi \frac{\partial^2}{\partial n^2} \bigl((1-n)S_n \bigr)\Bigr]^{-1/2}\bigg|_{n_0}= \frac{1}{\lambda^{n+1}} e^{(1-n)S_{n}} \Bigl[ 2\pi \frac{C_{E}(n)}{n^{2}}\Bigr]^{-1/2}\bigg|_{n_0}.	
\end{equation}
This shows a relation between the capacity of entanglement and the entanglement spectrum at the vicinity of the saddle point.

\section{Generating the potential and a byproduct}
\label{sec:gen}
\subsection{Generating $V(\phi)$ from the $V=0$ solution}
\label{sec:Vphi}

We will give a way to obtain the potential~\eqref{eq:V-alpha} and the solution~\eqref{eq:ansatz-d}--\eqref{eq:soln-d} \textit{without any manual input of unknown functions}. As a consequence, the potential~\eqref{eq:V-alpha} can be generated from the $V=0$ solution that was found in \cite{Horowitz:1991cd}.

It was known that the potential can be generated from a given function in the metric or the scalar field, and this method was used to construct many scalar potentials \cite{Gubser:2008ny,Li:2011hp,Cai:2012xh,Anabalon:2012ta,Feng:2013tza}. Generally, all parameters in the metric will enter the potential by this procedure. On the one hand, the parameters in the potential are model parameters, i.e., parameters in the Lagrangian. On the other hand, the solution parameters have integration constants such as mass and chemical potential that are not expected in the potential \cite{Feng:2013tza}. Therefore, it is nontrivial for these solution parameters to not be in the potential. In other words, requiring a parameter in the solution to not be in the potential gives a constraint on the potential. An observation in \cite{Ren:2019lgw} is that if we choose this parameter to be the spatial curvature $k$ with reasonable assumptions, the potential will be completely determined. We generalize the appendix A of \cite{Ren:2019lgw} to arbitrary dimensions in the following.

We consider the neutral solution for simplicity. The action is
\begin{equation}
S=\int d^{d+1}x\,\sqrt{-g}\left(R-\frac{1}{2}(\partial\phi)^2-V(\phi)\right),\label{eq:actiond-ES}
\end{equation}
where $V(\phi)$ is the potential of the scalar field $\phi$. We consider the following metric ansatz:
\begin{equation}
ds^2=e^{2\mathcal{A}(\bar{r})}(-h(\bar{r})dt^2+d\tsp\Sigma_{d-1,k}^2)+\frac{e^{2\mathcal{B}(\bar{r})}}{h(\bar{r})}d\bar{r}^2\,,\label{eq:ansatz-gubser}
\end{equation}
where $\bar{r}$ is the AdS radial coordinate, and the metric for the $(d-1)$-dimensional sphere, plane, and hyperbolic space can be written as
\begin{equation}
d\Sigma_{d-1,k}^2=\frac{dx^2}{1-kx^2}+x^2d\Omega_{d-2}^2\,,
\end{equation}
where $d\Omega_{d-2}^2$ is a $(d-2)$-dimensional sphere of unit radius. There is one gauge degree of freedom in the unknown functions $\mathcal{A}(\bar{r})$, $\mathcal{B}(\bar{r})$, $h(\bar{r})$, and $\phi(\bar{r})$, and it will be fixed by $\phi=\bar{r}$.

Equations of motion are obtained by the action~\eqref{eq:actiond-ES} with the metric ansatz~\eqref{eq:ansatz-gubser}. The Einstein's equation gives
\begin{gather}
\mathcal{A}'\mathcal{B}'=\frac{1}{2(d-2)}\phi'^2+\mathcal{A}''\,,\label{eq:B}\\
(e^{d\mathcal{A}-\mathcal{B}}h')'+2(d-2)e^{(d-2)\mathcal{A}+\mathcal{B}}k=0\,.\label{eq:h}
\end{gather}
The first equation comes from $G_{tt}$ and $G_{\bar{r}\bar{r}}$, and the second equation comes from $G_{\bar{r}\bar{r}}$ and $G_{xx}$. Solving the potential from $G_{\bar{r}\bar{r}}$ gives
\begin{equation}
V=\bigl[\frac{1}{2}\phi'^2h-d(d-1)\mathcal{A}'^2h-(d-1)\mathcal{A}'h'\bigr]e^{-2\mathcal{B}}+(d-1)(d-2)e^{-2\mathcal{A}}k\,.\label{eq:V}
\end{equation}
Other equations can be derived from~\eqref{eq:B}, \eqref{eq:h}, and \eqref{eq:V}. Starting with a given $\mathcal{A}(\bar{r})$, we can obtain $V(\bar{r})$ in the following way:
\begin{equation}
\mathcal{A}\xrightarrow{\eqref{eq:B}} \mathcal{B}\xrightarrow{\eqref{eq:h}} h\xrightarrow{\eqref{eq:V}} V\,.\label{eq:ABhV}
\end{equation}
The function $\mathcal{A}(\bar{r})$ plays the role of a generating function. Finally, replacing the function $V(\bar{r})$ with $V(\phi)$ gives the potential. A caveat is that only careful choices of $\mathcal{A}$ can we obtain a relatively simple $V(\phi)$.

The potential $V$ solved by~\eqref{eq:ABhV} generally depends on $k$. We require that the potential $V$ is independent of $k$, to have fewer parameters than the solution. The terms dependent on $k$ must cancel:
\begin{equation}
V=V^{(0)}+V^{(k)},\qquad V^{(k)}=0\,.
\end{equation}
We need an additional constraint: the function $h$ depends on $k$, and other functions are independent of $k$. In other words, for a given potential $V$, the only difference between the $k=0$ solution and the $k\neq 0$ solution is some terms $h^{(k)}$ in $h$. The motivation for this constraint is that if it is satisfied, the equation \eqref{eq:h} for $h$ is a linear equation, which we can take advantage of. The following will be based on the above constraints.

We decompose $h$ into a $k$-independent part and a $k$-dependent part: $h(\bar{r})=h^{(0)}+h^{(k)}$, where $h^{(0)}$ is the solution of $h$ at $k=0$. Similarly, we decompose the equation of motion into a part at $k=0$ and a part dependent on $k$. The part at $k=0$ requires that $\mathcal{A}$, $\mathcal{B}$, and $h^{(0)}$ satisfy the equations of motion with $V=V^{(0)}$, and the part dependent on $k$ requires that $h^{(k)}$ satisfy the equations of motion with $V=0$. As a consequence, the $V=V^{(0)}$, $k=0$ solution and the $V=0$, $k=1$ solution share the same generating function $\mathcal{A}$ (as well as $\mathcal{B}$ and $\phi$). Start with a solution with $V=0$ and $k=1$ as a seed, and we can use the procedure~\eqref{eq:ABhV} with $h=1$ and $k=0$ to obtain the potential $V(\phi)$.

The solution of $h$ from~\eqref{eq:h} is given by
\begin{equation}
h=\int e^{-d\mathcal{A}+\mathcal{B}}\left(-2(d-2)k\int e^{(d-2)\mathcal{A}+\mathcal{B}}d\bar{r}+C_2\right)d\bar{r}+C_1\,,
\label{eq:sol-h}
\end{equation}
where $C_1=C_2=0$ gives the solution for the system with $V=0$. The general $V(\phi)$ solution will contain the two integration constants coming from the second-order linear equation for $h$. If we take $C_1=1$ and $C_2\propto k$, we can obtain the potential $V(\phi)$ as~\eqref{eq:V-alpha}.

Now it boils down to solving the system with $V=0$. This is nontrivial, but has been achieved in \cite{Horowitz:1991cd}. (The AdS$_4$ case was in~\cite{Garfinkle:1990qj} earlier.) We shall not repeat the details. However, we use a more convenient coordinate system to show that the equations of motion are solvable when $V=0$. The metric ans\"atz is
\begin{equation}
ds^2=-e^{2A}hdt^2+e^{2B}\biggl(\frac{dr^2}{h}+r^2d\Sigma_{d-1,k}^2\biggr)\,,\label{eq:ansatz-lu}
\end{equation}
with the gauge \cite{Duff:1996hp}
\begin{equation}
A+(d-2)B=0\,.
\end{equation}
The independent equations are
\begin{align}
    &\phi'^2+2(d-1)\biggl(B''+(d-2)B'^2+\frac{d-1}{r}B'\biggr)=0\,,\\
    &\text{\small $h''-\biggl(2(d-1)-\frac{d-3}{r}\biggr)h'-2\biggl((d-1)B''+\frac{(d-1)^2}{r}B'+\frac{(d-2)}{r^2}\biggr)h+\frac{2(d-2)k}{r^2}=0$}\,,\\
    &V=-e^{-2B}\biggl(h''+\frac{3d-5}{r}h'+\frac{2(d-2)^2}{r^2}h-\frac{2(d-2)^2}{r^2}k\biggr).
\end{align}
The last equation shows that the function $h$ is easily solvable when $V=0$. Then $B$ can be solved by the second equation, and $A$ by the first equation.

Once we have obtained the metric, by converting from the metric \eqref{eq:ansatz-lu} to \eqref{eq:ansatz-gubser}, we obtain the generating function
\begin{equation}
e^\mathcal{A}=b\left(e^{\frac{d-2}{(d-1)\alpha}\bar{r}}-e^{-\frac{\alpha}{2}\bar{r}}\right)^{-\frac{1}{d-2}}.
\end{equation}
Alternatively, we can also use the solution of $\phi(r)$ as the generating function. Let $V_\alpha(\phi)$ be the potential~\eqref{eq:V-alpha} with parameter $\alpha$, a more general potential than \eqref{eq:V-alpha} is
\begin{equation}
V(\phi)=V_\alpha(\phi)+V_\text{extra},
\label{eq:V-general}
\end{equation}
where $V_\text{extra}$ comes from a nonzero $C_2$ in \eqref{eq:sol-h} at $k=0$. This potential has already been obtained in \cite{Feng:2013tza} by treating the solution of $\phi(r)$ as an ans\"atz. Here, we treat the $V=0$ solution as a seed to generate the general potential, and emphasize its naturalness. The extra terms in \eqref{eq:V-general} involve hypergeometric functions and look cumbersome. However, in the AdS$_4$ case, a six-exponential potential with a simple structure can be obtained as \eqref{eq:6-exp-V} below.

\subsection{Two neutral solutions with scalar hair in AdS$_4$}
\label{sec:two-neutral}
It was observed in \cite{Ren:2019lgw} that there are two neutral limits of the hyperbolic black holes~~\eqref{eq:ansatz-d}--\eqref{eq:soln-d}: a trivial neutral limit $b=0$ where we obtain the hyperbolic Schwarzschild-AdS black hole, and a nontrivial neutral limit $c=0$ where we obtain a hairy black hole. As a consequence, the hyperbolic black hole spontaneously develops a scalar hair below a critical temperature. This was used to analytically study phase transitions of the R\'enyi entropy \cite{Bai:2022obp}. In the following, we obtain a different neutral hyperbolic black hole with scalar hair for the same system~\eqref{eq:actiond} with \eqref{eq:V-alpha}.

Start with a more general action
\begin{equation}
S=\int d^{4}x\sqrt{-g}\biggl(R-\frac{1}{4}e^{-\alpha\phi}(F^1)^2-\frac{1}{4}e^{\phi/\alpha}(F^2)^2-\frac12(\partial\phi)^2-V(\phi)\biggr).\label{eq:action4}
\end{equation}
Let $V_\alpha(\phi)$ be the three-exponential potential~\eqref{eq:V-alpha}, and we consider the following six-exponential potential \cite{Ren:2019lgw}\footnote{Earlier works on deriving a six-exponential potential were \cite{Anabalon:2012ta,Feng:2013tza}, and various properties were studied in \cite{Anabalon:2019tcy,Anabalon:2017yhv,Anabalon:2020pez}, for example.}
\begin{equation}
V(\phi)=(1+\beta)V_\alpha(\phi)-\beta V_{-\alpha}(\phi)\,.
\label{eq:6-exp-V}
\end{equation}
The solution of the metric $g_{\mu\nu}$, gauge fields $A_\mu^{1,2}$, and dilaton field $\phi$ is
\begin{align}
& ds^2 =-f(r)\tsp dt^2+\frac{1}{f(r)}\tsp dr^2+U(r)\tsp d\Sigma_{2,k}^2\,,\label{eq:sol4}\\
& A^1 =2\gamma\sqrt{\frac{bc}{1+\alpha^2}}\left(\frac{1}{r_h}-\frac{1}{r}\right)dt\,,\\
& A^2 =2\alpha\sqrt{\frac{(\gamma^2-1)bc}{1+\alpha^2}}\left(\frac{1}{r_h-b}-\frac{1}{r-b}\right)dt\,,\\
& e^{\alpha\phi}=\left(1-\frac{b}{r}\right)^\frac{2\alpha^2}{1+\alpha^2},
\end{align}
with
\begin{align}
f &=\left(k-\frac{c}{r}\right)\left(1-\frac{b}{r}\right)^\frac{1-\alpha^2}{1+\alpha^2}+(1+\beta)\frac{r^2}{L^2}\biggl(1-\frac{b}{r}\biggr)^\frac{2\alpha^2}{1+\alpha^2}+(\gamma^2-1)\frac{bc}{r^2}\biggl(1-\frac{b}{r}\biggr)^{-\frac{2\alpha^2}{1+\alpha^2}}\nonumber\\
&\hspace{0.15\textwidth} -\beta\frac{r^2}{L^2}\biggl(1+\frac{1-3\alpha^2}{1+\alpha^2}\,\frac{b}{r}+\frac{(1-\alpha^2)(1-3\alpha^2)}{(1+\alpha^2)^2}\,\frac{b^2}{r^2}\biggr)\biggl(1-\frac{b}{r}\biggr)^{\frac{1-\alpha ^2}{1+\alpha ^2}},\label{eq:fsol4}\\
U &=r^2\biggl(1-\frac{b}{r}\biggr)^\frac{2\alpha^2}{1+\alpha^2}.\label{eq:Usol4}
\end{align}
The solution has parameters $b$, $c$ and $\gamma$ in addition to $\alpha$ and $\beta$. By taking $\gamma=1$, we obtain an EMD system with a six-exponential potential of the dilaton. The curvature singularity is at $r=0$ and $r=b$. The mass is given by
\begin{equation}
M=\frac{V_\Sigma}{8\pi G}\biggl(c+k\frac{1-\alpha^2}{1+\alpha^2}b-\beta\frac{(1-\alpha^2)(1-3\alpha^2)(3-\alpha^2)}{2(1+\alpha^2)^3L^2}b^3\biggr).
\end{equation}
The trivial neutral limit of this solution is $b=0$, where the scalar field $\phi$ vanishes, and the potential becomes the cosmological constant. We are interested in the nontrivial neutral limit $c=0$ later.

In \cite{Anabalon:2012ta,Feng:2013tza} and \cite{Faedo:2015jqa}, it was found that there are two analytic solutions to the same system with potential $V_\alpha(\phi)$, and this was explained in \cite{Nozawa:2022upa} (see also \cite{Anabalon:2017yhv}). Here we give a simple explanation as follows. The six-exponential potential has an additional parameter $\beta$. By taking $\beta=0$, we obtain a solution for the three-exponential potential $V_\alpha(\phi)$; by taking $\beta=-1$, $\alpha\to-\alpha$ and $\phi\to-\phi$, we obtain a different solution for the same potential. Thus, the nontrivial neutral limit $c=0$ gives two different hyperbolic black holes with scalar hair. One neutral solution was studied in \cite{Ren:2019lgw,Bai:2022obp}, and the other one is given by
\begin{align}
f(r) &=\biggl[-1+\frac{r^2}{L^2}\biggl(1+\frac{1-3\alpha^2}{1+\alpha^2}\,\frac{b}{r}+\frac{(1-\alpha^2)(1-3\alpha^2)}{(1+\alpha^2)^2}\,\frac{b^2}{r^2}\biggr)\biggr]\biggl(1-\frac{b}{r}\biggr)^{\frac{1-\alpha ^2}{1+\alpha ^2}},\\
\phi &=-\frac{2\alpha}{1+\alpha^2}\ln\left(1-\frac{b}{r}\right).
\end{align}
The boundary condition for the scalar field corresponds to a multi-trace deformation in the dual CFT \cite{Witten:2001ua}. The boundary conditions corresponding to the triple-trace deformation are different for the two solutions, while the boundary conditions corresponding to the double-trace deformation are the same for the two solutions.

\section{Summary and discussion}
\label{sec:sum}
In this paper, we have calculated the supersymmetric R\'{e}nyi entropies (with a spherical entangling surface) from a class of hyperbolic black holes with scalar hair. Our findings are summarized as follows:

\begin{itemize}
\item By employing a class of hyperbolic black holes with scalar hair, we have explicitly obtained the holographic supersymmetric R\'{e}nyi entropies. Our results not only corroborate many established outcomes, but also introduce additional findings with distinctive properties.

\item We have calculated the supersymmetric capacity of entanglement and showed that it cannot be mapped to the heat capacity of hyperbolic black holes due to the fact that the BPS condition gives a constraint between the temperature and the chemical potential.

\item From the R\'{e}nyi entropies that are analytic at $n=\infty$, we have calculated the entanglement spectrum as convolutions of generalized hypergeometric functions.

\item We have shown that the potential of the EMD system can be generated from a $V(\phi)=0$ solution.

\item There are two nontrivial neutral limits of the EMD system, giving hyperbolic black holes with scalar hair. Scalar condensation may happen at sufficiently low temperatures.
\end{itemize}

The following topics need further investigation: (i) The CFT calculations for the models in table~\ref{tab:1}. (ii) Violation of inequalities when $\alpha>\alpha_*$, especially the 1-charge black hole in AdS$_4$ as a top-down special case. (iii) The geometric interpretation of the supersymmetric R\'enyi entropy. (iv) Whether there are phase transitions for this class of hyperbolic black hole solutions.

\acknowledgments
J.R. thanks Xiaoxuan Bai, Antonio Gallerati, Song He, Masato Nozawa, Yi Pang, Chiara Toldo, and Yang Zhou for helpful communications. This work was supported in part by the NSF of China under Grant No. 11905298, and Fundamental Research Funds for the Central Universities, Sun Yat-sen University under Grant No. 23qnpy61.

\appendix
\section{Special cases of $D=4,5,6,7$ supergravities}
\label{sec:STU}
In STU supergravities, there are U$(1)^4$ gauge fields in AdS$_4$, U$(1)^3$ gauge fields in AdS$_5$, and U$(1)^2$ gauge fields in AdS$_7$ \cite{Cvetic:1999xp}. Special cases of them can be reduced to EMD systems. They are 1-charge, 2-charge, and 3-charge black holes in AdS$_4$; 1-charge and 2-charge black holes in AdS$_5$; and 1-charge black hole in AdS$_7$.

The AdS$_4$ Lagrangian is
\begin{equation}
\mathcal{L}=R-\frac{1}{2}(\partial\vec{\phi})^2+8g^2(\cosh\phi_1+\cosh\phi_2+\cosh\phi_3)-\frac{1}{4}\sum_{i=1}^4 e^{\vec{a}_i\cdot\vec{\phi}}(F^i_{(2)})^2\,,
\end{equation}
where $\vec{\phi}=(\phi_1, \phi_2, \phi_3)$, $\vec{a}_1=(1, 1, 1)$, $\vec{a}_2=(1, -1, -1)$, $\vec{a}_3=(-1, 1, -1)$, and $\vec{a}_4=(-1, -1, 1)$. More details can be found in \cite{Cvetic:1999xp}. The solution is given by \cite{Duff:1999gh,Sabra:1999ux}
\begin{align}
ds^2 &=-(H_1H_2H_3H_4)^{-1/2}fdt^2+(H_1H_2H_3H_4)^{1/2}(f^{-1}dr^2+r^2d\tsp\Sigma_{2,k}^2)\,,\\
X_i &=H_i^{-1}(H_1H_2H_3H_4)^{1/4}\,,\\
A_{(1)}^i &=\sqrt{k}(1-H_i^{-1})\coth\beta_i\tsp dt\,,
\end{align}
with $X_i=e^{-\frac{1}{2}\vec{a}_i\cdot\vec{\phi}}$, and
\begin{equation}
f=k-\frac{\mu}{r}+4g^2r^2(H_1H_2H_3H_4)\,,\qquad H_i=1+\frac{\mu\sinh^2\beta_i}{kr}\,.
\end{equation}

We call this general solution (1+1+1+1)-charge black hole, or 4-charge black hole in AdS$_4$ if there is no confusion. For special cases, the following naming convention is used in the literature.
\begin{center}
\renewcommand{\arraystretch}{1.2}
\setlength\doublerulesep{0.2pt}
\begin{tabular}{|c|c|}
\hline\hline
$H_i$ ($i=1,2,3,4$) & Name \\
\hline\hline
$H_1=H_2=H_3$, $H_4\quad (\neq 1)$ & (3+1)-charge black hole in AdS$_4$\\
\hline
$H_1=H_2$, $H_3=H_4\quad (\neq 1)$ & (2+2)-charge black hole in AdS$_4$\\
\hline
$H_1=H_2=H_3=H$, $H_4=1$ & 3-charge black hole in AdS$_4$\\
\hline
$H_1=H_2=H$, $H_3=H_4=1$ & 2-charge black hole in AdS$_4$\\
\hline
$H_1=H$, $H_2=H_3=H_4=1$ & 1-charge black hole in AdS$_4$\\
\hline
$H_1=H_2=H_3=H_4=H$ & RN-AdS$_4$ black hole\\
\hline\hline
\end{tabular}
\end{center}

The AdS$_5$ Lagrangian is
\begin{equation}
\mathcal{L}=R-\frac{1}{2}(\partial\vec{\varphi})^2+4g^2\sum_i X_i^{-1}-\frac{1}{4}\sum_{i=1}^4 X_i^{-2} (F^i_{(2)})^2\,.
\end{equation}
The solution is \cite{Behrndt:1998jd}
\begin{align}
& ds^2=-(H_1H_2H_3)^{-2/3}fdt^2+(H_1H_2H_3)^{1/3}(f^{-1}dr^2+r^2d\tsp\Sigma_{3,k}^2)\,,\\
& X_i=H_i^{-1}(H_1H_2H_3)^{1/3}\,,\\
& A_{(1)}^i=\sqrt{k}(1-H_i^{-1})\coth\beta_idt\,,
\end{align}
with
\begin{equation}
f=k-\frac{\mu}{r^2}+g^2r^2(H_1H_2H_3)\,,\qquad H_i=1+\frac{\mu\sinh^2\beta_i}{kr^2}\,.
\end{equation}
We call this general solution (1+1+1)-charge black hole, or 3-charge black hole in AdS$_5$ if there is no confusion. For special cases, we have (2+1)-charge black hole, 2-charge black hole, 1-charge black hole in AdS$_5$. When $H_1=H_2=H_3$, we have the RN-AdS$_5$ black hole.

The static AdS$_6$ black hole metric is \cite{Chow:2011fh,Hosseini:2019and}
\begin{align}
& ds^2= -\frac{9}{2}(H_1H_2)^{-3/4}f dt^2 + (H_1H_2)^{1/4}(f^{-1}dr^2+r^2 d\tsp\Sigma_{4,k}^2),,\\
&X_i= H_i^{-1}(H_{1}H_{2})^{3/8}\,,\\
&A_{(1)}^i =\sqrt{k}\coth\beta_i(1-H_i^{-1})dt\,,
\end{align}
with
\begin{equation} 
f(r) = k-\frac{\mu}{r^3}+\frac{2}{9}r^2H_1 H_2 \, , \qquad H_i = 1+\frac{\mu\sinh^2\beta_i}{kr^3} \, .
\end{equation}
Here, $i=1,2$. We call this solution (1+1)-charge black hole in AdS$_6$. Special cases are 2-charge black hole and 1-charge black hole in AdS$_6$. Note that the 2-charge case ($H_1=H_2$) is not the RN-AdS$_6$ black hole.

The AdS$_7$ Lagrangian is
\begin{equation}
e^{-1}\mathcal{L}=R-\frac{1}{2}(\partial\vec{\varphi})^2 - g^2 V -\frac{1}{4} \sum_{i=1}^2 e^{\vec{a_i}\cdot\vec{\varphi}}(F^i_{(2)})^2\,.
\end{equation}
The solution is \cite{Cvetic:1999xp}
\begin{align}
& ds_7^2 =-(H_{1}H_{2})^{-4/5} f dt^2 +(H_{1}H_{2})^{1/5} (f^{-1} d r^2 + r^2 d\Omega_{5,k}^2)\,,\\
&X_i=  H_i^{-1}(H_{1}H_{2})^{2/5}\,,\\
&A_{(1)}^i =\sqrt{k}\coth\beta_i(1-H_i^{-1})dt\,,
\end{align}
with
\begin{equation}
 f= k -\frac{\mu}{r^4} + \frac{1}{4} g^{2} r^{2} H_{1}H_{2}\,,\qquad  H_i = 1+\frac{\mu \sinh^2\beta_i}{kr^4}\ ,
\end{equation}
We call this solution (1+1)-charge black hole in AdS$_7$. Special cases are 2-charge black hole and 1-charge black hole in AdS$_7$. Note that the 2-charge case ($H_1=H_2$) is not the RN-AdS$_7$ black hole.

For EMD truncations of these solutions, we need to shift the radial coordinate $r\to r-\mu\sinh^2\beta$ to obtain the solutions \eqref{eq:ansatz-d}--\eqref{eq:soln-d}.

\section{Fayet-Iliopoulos gauged supergravity}
\label{sec:sugra}

We briefly review the $\mathcal{N}=2$, $D=4$ supergravity with Abelian Fayet-Iliopoulos (FI) gaugings that can be reduced to EMD systems. See \cite{Nozawa:2022upa} for more details and references. For $n_V$ number of abelian vector multiplets~\cite{Andrianopoli:1996cm},
the model describes $n_{V}+1$ vector fields $A_{\mu}^{I}$($I=0,1,\dots,n_{V}$) and $n_{s}=n_{V}$ complex scalars fields $z^{\alpha}(\alpha=1,\dots,n_{s})$.
These scalars parametrize an $n_{V}$-dimensional Hodge-K\"ahler manifold, which is the base of a symplectic bundle with covariantly holomorphic section
\begin{equation}
\mathcal{V}=\biggl(
\begin{array}{c}
X^I\\
F_I   
\end{array}
\biggr)\,, \qquad 
\mathcal{D}_{\bar \alpha}\mathcal{V}= \partial_{\bar \alpha}\mathcal{V}-\frac{1}{2} (\partial _{\bar \alpha}\mathcal{K})\mathcal{V}=0 \,, 
\end{equation}
where $\mathcal{V}$ obeys the symplectic constraint $\langle\mathcal{V}, \bar{\mathcal{V}}\rangle \equiv X^I\bar F_I-F_I\bar X^I= i$ and $\langle \mathcal{V}, \partial_\alpha {\mathcal{V}}\rangle=0$; $\mathcal{K}=\mathcal{K}(z^\alpha, \bar z^\alpha)$ is the K\"ahler potential, and $\mathcal{D}_\alpha$ denotes the K\"ahler covariant derivative. Writing
\begin{equation}
\mathcal{V}= e^{K/2}v,\qquad v= \biggl(
\begin{array}{c}
Z^I\\
\frac{\partial}{\partial Z^I}F(Z) 
\end{array}
\biggr),
\end{equation}
where $v$ is the holomorphic symplectic vector. In appropriate symplectic frame, we assume the existence of prepotential $F$ that is a homogeneous function of degree two.

The bosonic gauged Lagrangian is
\begin{equation}
\label{Lag0}
\mathcal{L}=\frac{1}{2} (R-2 V) \star 1 -g_{\alpha\bar \beta}dz^\alpha \wedge \star d\bar z^{\bar\beta} 
+\frac{1}{2}I_{IJ}F^I \wedge \star F^J +\frac{1}{2}R_{IJ}F^I \wedge  F^J \,.
\end {equation}
where the $n_{V}+1$ vector field strengths are $F^{I}=dA^{I}$; $I_{IJ}={\rm Im}\mathcal{N}_{IJ}$, $R_{IJ}={\rm Re}\mathcal{N}_{IJ}$, where $\mathcal{N}_{IJ}$ is defined by the relations
$F_{I}=\mathcal{N}_{IJ}X^{J}$ and $\mathcal{D}_{\bar\alpha}\bar F_I = \mathcal{N}_{IJ}\mathcal{D}_{\bar\alpha}\bar X^J$. The scalar potential is
\begin{equation}
V = -2g_{I}g_{J}\left(I^{IJ}+8\bar X^I X^J\right)\,, 
\end {equation}
where $I^{IJ}$ is the inverse of $I_{IJ}$, and $g_I$ is the FI coupling constants.

We consider the following prepotential of $\mathcal{N}=2$ supergravity with one complex scalar ($n_V=1$):
\begin{equation}
F(X)=-\frac{i}{4} (X^0)^n (X^1)^{2-n}\,.
\end{equation}
The values of the parameter $n=1, 1/2$, and $3/2$ correspond to special cases of STU supergravity. This is a truncation of the STU model with the prepotential
\begin{equation} 
F_{\rm STU}(X)=-\frac{i}{4}\sqrt{X^0X^1X^2X^3}\,.
\end{equation}
Setting $Z^0=1$ and $Z^1=z$, the symplectic vector is $v=(1, z, -\frac{i}{4} n z^{2-n}, -\frac{i}{4} (2-n) z^{1-n})^T$. The system \eqref{eq:action4} is obtained by further truncating the theory to a single real scalar $z=\bar z$ and the purely electrically charged case $F^I \wedge F^J=0$.

\section{The IR geometry for $\alpha=\alpha_*$}
\label{sec:IR}

In the special case $\alpha=\alpha_*:=(d-2)\sqrt{\frac{2}{d-1}}$, the black hole under the BPS condition has a distinctive IR geometry at zero temperature: it has a curvature singularity. For comparison, when $0\leq\alpha<\alpha_*$, the IR geometry at zero temperature is AdS$_2\times\mathbb{H}^{d-1}$, i.e., it has a degenerate horizon.

When $\alpha=\alpha_*$, the IR limit of the geometry at zero temperature is
\begin{equation}
ds^2=(r-1)^\frac{2}{d-1}\biggl(-(r-1)dt^2+\frac{dr^3}{(r-1)^3}+d\Sigma_{d-1}^2\biggr).
\end{equation}
By the change of variables
\begin{equation}
\tilde{r}=(r-1)^{-1/2},
\end{equation}
the IR geometry is written as
\begin{equation}
ds^2=\tilde{r}^{\frac{2\theta}{d-1}}\biggl(-\frac{dt^2}{\tilde{r}^{2\mathsf{z}}}+\frac{d\tilde{r}^2+d\Sigma_{d-1}^2}{\tilde{r}^2}\biggr),\label{eq:ztheta1}
\end{equation}
which is a hyperscaling-violating geometry with the spatial part being $\mathbb{H}^{d-1}$. The Lifshitz scaling exponent $\mathsf{z}$ and the hyperscaling violation exponent $\theta$ are
\begin{equation}
\mathsf{z}=2,\qquad \theta=d-3.
\end{equation}
In particular, when $d=3$, it is a Lifshitz geometry with the spatial part being $\mathbb{H}^{d-1}$.

It is interesting to make a comparison to the planar black hole solutions of the same EMD system~\eqref{eq:actiond}. The IR geometries of extremal planar black holes are \cite{Ren:2019lgw}:
\begin{itemize}
\item $0<\alpha<(d-2)\sqrt{\frac{2}{d(d-1)}}$. The IR geometry is AdS$_2\times\mathbb{R}^{d-1}$.
\item $\alpha=(d-2)\sqrt{\frac{2}{d(d-1)}}$. The IR geometry is conformal to AdS$_2\times\mathbb{R}^{d-1}$ \cite{Gubser:2012yb,Gouteraux:2014hca}.
\item $\alpha>(d-2)\sqrt{\frac{2}{d(d-1)}}$. The extremal limit of the EMD system~\eqref{eq:actiond} is the same as an Einstein-scalar system. The IR geometry is a hyperscaling-violating geometry.
\end{itemize}

\end{document}